# Permissioned Blockchain Technologies for Academic Publishing

Petr Novotny, Qi Zhang, Richard Hull, Salman Baset, Jim Laredo, Roman Vaculin, Daniel L. Ford, Donna N. Dillenberger

*IBM T.J. Watson Research Centre*

**Abstract.** Academic publishing is continuously evolving with the gradual adoption of new technologies. Blockchain is a new technology that promises to change how individuals and organizations interact across various boundaries. The adoption of blockchains is beginning to transform diverse industries such as finance, supply chain, international trade, as well as energy and resource management and many others. Through trust, data immutability, decentralized distribution of data, and facilitation of collaboration without the need for centralized management and authority, blockchains have the potential to transform the academic publishing domain and to address some of the current problems such as productivity and reputation management, predatory publishing, transparent peer-review processes and many others. In this paper, we outline the technologies available in the domain of permissioned blockchains with focus on Hyperledger Fabric and discuss how they can be leveraged in the domain of academic publishing.

## 1. Introduction

Blockchain technologies are increasingly adopted in various industries [1-6], such as finance, supply chain, international trade, energy and resource management. Blockchains facilitate a direct interaction and data exchange between participants spanning organizational, national, regulatory and other boundaries in a trusted and reliable manner. The use of blockchains provides trust, transparency and facilitates transformation of established processes and forms of interaction.

Academic publishing focused on the distribution of academic research and scholarship, is centered on information processing. It involves a range of processes including acquisition, peer-review, editorial quality assurance, and many others, involving participants distributed across the globe. The interactions between participants are mostly facilitated by the use of disconnected and frequently transparency lacking technologies such as internet-based submission management systems, emails, various publishing websites, digital libraries, and social networking services. The introduction of blockchains into this ecosystem is expected to facilitate transparency and disrupt existing research processes, content discovery, and access to the scientific research among others. Moreover, blockchains have the potential of a wider evolution of research and academic publishing through the transformation of the established processes and relationships between the involved participants. The application of blockchain in this domain is becoming actively researched area [44, 45].

Blockchain is a system of record based on a distributed immutable ledger that is shared, replicated, and continuously synchronized among peers (i.e.: software agents) hosted by participants of a decentralized network. The distributed ledger records transactions, such as an exchange of documents or data, recorded by the participants of the network. The decentralized network typically does not have a single governing authority. Instead, depending on the scenario, it may be governed by a consortium of participants or democratically by stakeholders of the network. Blockchain allows to integrate business processes, systems, and whole ecosystems and to establish a reliable source of truth. In blockchain ledger, every record is timestamped and has a cryptographic signature, thus making the ledger an auditable and immutable history of all transactions in the network. Moreover, through decentralization, blockchains facilitate wide distribution and availability of data.

The introduction of blockchains into the academic publishing ecosystem is expected to address the following issues as well as to facilitate deeper collaboration between the researchers, publishers, funding organizations and other actors.

- Trusted management of published materials
- Integration of sources of published materials and data
- Trusted citing and referencing
- Productivity and citation metrics
- Reputation, performance and information about individual researches, institutions, laboratories, departments and teams
- Global wide reputability and trust of sources of information and data
- Identification and elimination of predatory journals, conferences and other types of channels
- A transparent peer-review processes
- Opportunity for reviewer feedback
- Intellectual property management, including open and paid access

Blockchains can be categorized as either public and permissionless such as Bitcoin [7] and Ethereum [8], or private and permissioned such as Hyperledger Fabric [9, 10] and Corda [11]. Each of these types provides different features and capabilities for the design of systems based on these platforms. In this paper, we focus on the distinctive features of private blockchains and of the technologies based on these blockchains. We highlight the relevant capabilities and identify their potential for use in the academic publishing domain.

## 2. Overview of Blockchain Technologies

After the initial successful emergence of the first well-known blockchain platform Bitcoin [7] providing a universal crypto-currency, several many other blockchain based cryptocurrencies quickly followed. At the same time started to emerge more general blockchain platforms such as Ethereum [8], which offer broad-spectrum data processing and exchange capabilities between participants. These networks are open to anyone to join and participate and typically use some form of reward to motivate participants to provide infrastructure and resources to operate the network. These "public" blockchain networks are designed to be used in a wide range of applications utilizing the globally distributed infrastructure. The applications based on public blockchains can leverage the wide availability, free access, and storage of the networks highly available network and computation resources, and as well as relatively inexpensive processing of transactions. The drawbacks include the amount of resources necessary to operate the networks needed to provide a reliable global operation. For example, currently the operation of Bitcoin consumes as much energy as the entire country of Austria [12] and the energy consumption increases with the continuous expansion of the network. Additionally, public blockchain networks typically have a low throughput of transactions i.e. about 7 and 15 transactions per second in Bitcoin [13] and Ethereum [14] networks respectively. An important architectural consideration is the openness and anonymity of these networks, which provide only limited support of privacy and security. Therefore, they are not suitable for applications which require a high degree of security.

In parallel to the public blockchains, started to emerge the private blockchains, which include platforms such as Hyperledger Fabric [9, 10] and Corda [11]. These private blockchain platforms provide closed environments, which only permitted participants can join. The networks provide access control mechanisms, which allow to constrain access to data and actions of the participants. Unlike in public blockchains where participants are typically anonymous to each other, in private blockchains, every participant has a well-known identity associated with all records created by that participant. The recording of identity (i.e. signing of records) allows to establish non-repudiatable history of records by providing evidence of origin as well as approval of transactions recorded into the immutable shared ledger.

In its core, a blockchain network consists of a series of peers (i.e.: software agents), which together form the blockchain network. Each peer of the network maintains a copy of an immutable ledger. The ledger consists of a series of ordered transactions organized into cryptographically connected blocks. The cryptographic connection allows to verify that the ledger was not tampered with and that none of the transactions was modified since it was recorded. The transactions are generated by a smart contract, a business logic component (i.e.: a software code), which is agreed upon by the participants of the blockchain network. An identical copy of the smart contract is hosted on the peers of the blockchain network. When client submits a request for transaction into the network, the peers invoke the smart contract and calculate the transaction. To achieve a consistent and reliable state across the blockchain network, the blockchain uses a consensus mechanism, which prescribes the conditions of an agreement between the peers on the outcome of the transaction. Only if the outcome of invocation of the smart contract by the prescribed number of peers yields the same outcome, the transaction is considered valid and can be committed into the ledger. This mechanism prevents faulty or corrupted peers from introducing invalid transactions into the ledger. The valid transaction is distributed between the peers to maintain a consistent and up to date state of the ledger of each peer of the network.

Among the key differences between public and private blockchains are the mechanisms of consensus. Public blockchains utilize the proof of work [7] or proof of stake [8] in which a decisive size or weight of the network must agree on a transaction before it is accepted into the shared ledger. These mechanisms work well with a large size of the blockchain network. However, small networks may be vulnerable to attacks which can force inclusion of fraudulent transactions into the ledger [15]. Moreover, due to the inherent complexity of these mechanisms, the transaction processing in the public networks tend to be slow (e.g.: Bitcoin and Ethereum process few transactions per second). Private blockchains, on the other hand, provide the ability to define custom consensus policies tailored to the specific use case. Hence, every application can have its own definition of who must verify and agree to a transaction before it can be included in the shared ledger. In general, these mechanisms are fast and yield significantly better performing blockchain networks. Combined with an optimized software and hardware [16], private blockchain networks can provide significantly better performance (e.g.: Hyperledger Fabric can process thousands of transactions per second [17]).

The blockchain technologies are quickly evolving. For example, the Hyperledger Fabric project has currently more than 250 member companies and large number of software engineers actively participating on development of new features. New technologies such as new methods of smart contracts [23] or increasing reliability [43] are proposed and implemented with every new version. The selection of appropriate blockchain platform for a new system is thus a challenging task which requires understanding of number of areas including the system design, security topics such as cryptography and privacy, as well as the specific attributes of the blockchain platforms.

## 3. Consensus in Blockchain

In this section, we look at the consensus mechanism in context of trust. In blockchain, consensus [18] is the process of agreeing on what the ledger's "true" contents are. The consensus mechanism forms the foundation of the trust one places in a blockchain based ledger. In its simplest form, forming a consensus simply means that the set of computers maintaining a Blockchain agree that any particular block in the chain contains data that is a semantic component of the ledger, and that such a block is correctly ordered/positioned in the Blockchain. The latter point is important because block order corresponds directly to temporal order, which, if the block data represents financial transactions, for instance, has direct monetary consequences if wrong.

It is important to note that the process of forming a consensus does not, in general, endorse the semantic validity of the data in the blocks added to a Blockchain. For example, one could store a string in a block with the sentence "The Moon is made of green cheese." A consensus could be produced by the computers maintaining the Blockchain such that they all agree that a block containing the sentence is part of the Blockchain, but they can't decide if the sentence is actually true. A later observer of the Blockchain can trust that the sentence was the one that was intended to be preserved in the Blockchain, and that it hasn't been changed since.

On the surface, forming a consensus seems like it might be a simple problem, and in a perfect world, it might be, but in the real world many things can go wrong. The main challenge is protecting the process in the face of ongoing disruptions, including overt attempts by nefarious actors attempting to alter the ledger to their benefit, or to simply delay or disrupt the consensus process itself. Hackers, for instance, might compromise one or more of the machines maintaining the blockchain and send out contradictory or confusing communications to the others. Other disruptions include transmission delays, equipment failures (machines do crash, and the power does fail), and software bugs. Of course, any and all of these things could help at once, and the process still needs to produce a consensus, if it can, and it needs to do so in a timely manner.

One of the simplest consensus processes is the one used for the blockchain underlying the cryptocurrency Bitcoin. That blockchain is subject to developing branches leading to alternative collections of blocks to be (exclusively) interpreted to contain parts of the ledger, or not. The simple solution for Bitcoin is to define the longest branch to contain blocks of the "true" ledger, any other branch does not. This process is simple to understand and implement, with the computers maintaining the blockchain programmed to extend the longest branch, thus reinforcing the consensus choice.

There are various consensus schemes based on having the computers vote on the consensus. This approach typically appoints one of the computers to be a "leader" that is responsible for tabulating the votes. The process is neither simple to understand nor to implement, mostly because it needs to address a number of complex issues in the face of failure and malicious attack (e.g., leader selection). These issues, the overhead of exchanging votes, and the need for other intermachine communications that are in addition to the basic exchange of block data, tends to limit performance.

## 4. Blockchain Security Considerations

The use of blockchain requires careful consideration of the security attributes and risks. Blockchains are often public key infrastructure (PKI) systems at scale. Blockchain derives its immutable properties from the underlying cryptographic primitives. These cryptographic primitives include asymmetric encryption (based on ECC or RSA) and digital signatures. The responsibility for managing the keys associated with digital signatures is on the "users" or "applications" that interact with blockchain.

The distributed nature of blockchain means that the storage of data on different nodes is governed by a consensus algorithm – in context of security, the process by which nodes agree that a data that is cryptographically immutable can be stored within their ledger and that other nodes also store the same data in their ledgers. Public blockchains are vulnerable to attacks based on creating network partitions, or by exploiting certain properties of consensus algorithms (e.g., creating server farms to solve cryptographic puzzles or launching denial of service attacks).

Another vulnerability of public blockchains is linked to the use of private keys to secure and store data. The use of private keys was exploited by several well-known attacks on cryptocurrencies [19] and similar key centered types of attacks.

In general, the risks in public blockchains are high because the participants involved in the blockchain are unknown to each other. Consequently, the blockchain must provide strong guarantees (in terms of algorithms, and rewards) to prevent malicious actors from taking over blockchain. In contrast, in private blockchains, the actors are known and constrained by access control mechanisms, which eliminate many of the attack vectors. However, both types of blockchains run smart contracts, which are computer code. Without appropriate safeguards, such as input validation or preventing the code from running indefinitely, the smart contracts can cause significant harm to the underlying blockchain [20].

## 5. Blockchain-as-a-Platform

Hyperledger Fabric is a permissioned and enterprise level blockchain technology. Thanks to its advantages, such as enhanced on-chain data confidentiality, high transaction throughput, and low latency, enterprises and research institutions are increasingly interested to innovate their businesses processes and operational models with the use of Fabric [21]. However, in many scenarios, operating a high-performance Fabric network on on-premises can be an expensive proposition. Aside from the cost of the hardware, it may be necessary to hire blockchain experts to set up the network, configure it to the best performance, and develop the smart contracts. To solve this challenge, running the blockchain network in the cloud is a model that enables an easy access to blockchain technologies. With a cloud-based blockchain platform, customers can focus more on innovating their businesses and operations using the blockchain, while the cloud service provider takes the responsibility of delivering a high-quality blockchain platform. As one of the pioneers in this area, IBM develops IBM Blockchain Platform (IBP) [22] a full stack blockchain-as-a-service (BaaS) offering with features including high performance, enhanced security, and high availability, available in a globally distributed public cloud.

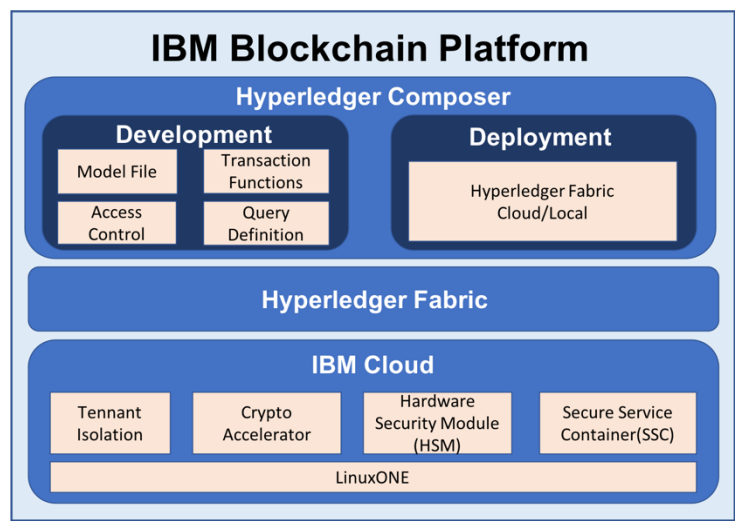

**Figure 1.** The architecture of IBM Blockchain Platform.

Figure 1 depicts the architecture of IBP, which consists of Hyperledger Fabric - the blockchain network service, and Hyperledger Composer – the blockchain application development framework, which aims at accelerating the time it takes to develop blockchain applications. Instead of writing the smart contracts from scratch, Hyperledger Composer provides a convenient layer and business-level abstractions to help customers create a blockchain network, implement smart contracts, and deploy them on Fabric. The platform also provides web-based tools for customers to manage and monitor their blockchain networks.

By running the business processes on a cloud-based blockchain platform such as IBP, the two primary concerns are the performance and data privacy. A good performance allows the platform to handle a high frequency of requests (e.g.: conference registrations, paper/review submissions) at peak times, while the secure environment guarantees that the sensitive data, for example, the scholars' personal information or unpublished work, will not be leaked out. To address these concerns, IBP uses the LinuxONE mainframes [23] as the infrastructure of the blockchain platform. Aside from the high level of isolation among different tenants, LinuxONE provides many other advantages, such as high performance and data privacy among others. To achieve thousands of transactions per second on IBP, LinuxONE uses crypto accelerators, which can accelerate the cryptographic functions that are extensively used in Fabric. LinuxONE provides hardware level virtualization and isolation mechanism Secure Service Container (SSC) yielding high security including the prevention of insider attacks [24]. This category of attacks is one of the major vulnerabilities in cloud services [25], in which the data is leaked by the illegal access of the insiders of the cloud services, such as the platform administrators. SSC prevents the insider attack by automatically encrypting all the code and data in memory and on the drive at all times. With SSC, even the platform administrators do not have the decryption keys and thus cannot access the data. Another critical component in IBP to preserve data privacy is the Hardware Security Module (HSM) which holds the decryption keys of the data in SSCs. The decryption keys never leave HSM as well as cannot be accessed from outside of HSM. This mechanism guarantees that even if someone successfully makes a copy of the SSC data, it will not be possible to decrypt the data.

BaaS simplifies and accelerates the adoption of the blockchain technologies for many industries including academic publishing. Currently, the large numbers of academic publishing processes, publication and citation information, conference or journal access sites, and many other activities are managed on various disconnected platforms. BaaS can help to effectively integrate the academic publishing domain. We envision a BaaS-based academic publishing ecosystem that can be accessed by all the participants and where all the processes and activities such as peer review and citation metrics can be recorded with high credibility. This ecosystem will not only bring more efficient management and highly trusted governance of the academic data but also promote the collaboration among the scholars and publishers by effectively sharing information.

## 6. Secure Blockchain Analytics

One of the benefits of using blockchain in academic publishing is that large amount of trusted data related to various aspects of academic publishing processes, such as peer reviews and citations, can be accumulated on the shared ledger of blockchain. Analytics tools suitable for blockchain [41] will allow to extract the real value out of such data beyond just collecting the transactions. For instance, blockchain-enabled analytics tools need to be able to seamlessly operate on blockchain data which has the form of key-value pairs in the case of Hyperledger Fabric and other platforms. Additionally, the same security model and confidentiality constraints need to be followed if security and trust are of importance. Ideally, it is desirable that the notion of trust, enabled by blockchain extends to the blockchain analytics as well.

The blockchain-enabled trusted analytics can have many forms ranging from standard descriptive analytics to predictive or machine learning based tools. Similarly, architectures and implementations can vary significantly. Analytics tools can be tightly integrated with blockchain platforms operating directly on the blockchain data without extracting it to any external system, or the blockchain data can be securely extracted to external analytics platforms by providing appropriate data connectors. Analytics methods can be implemented in the smart contract (on-chain analytics), in the blockchain client applications, or completely externally. Specific architectural choices depend on concerns such as the required level of security, needs to operate in or near real-time, constraints for moving data, need to integrate with external off-chain datasets, etc.

In this section, we illustrate some aspects of blockchain-enabled secure analytics by briefly describing an analytics tool that we have developed for blockchain solutions leveraging Hyperledger Fabric. The tool provides a descriptive analytics service tightly integrated with the Hyperledger Fabric platform. The service uses the same security model and authentication as Hyperledger Fabric and it can directly access the blockchain data without the need to Extract Transform and Load ETL outside of the platform. We designed the service in a solution-independent way so that it can be used with any blockchain solution.

To the user, the blockchain analytics service is exposed as a web-based configurable analytics dashboard. It allows users (1) to provide blockchain authentication credentials, (2) to create and manage dynamic descriptive queries, such as time-series aggregations, spatiotemporal and top-N analytics, as reusable and embeddable widgets, and (3) to visualize the results of the descriptive queries as charts in widgets and dashboards. Figure 2 shows a sample dashboard we configured for an existing supply chain blockchain solution.

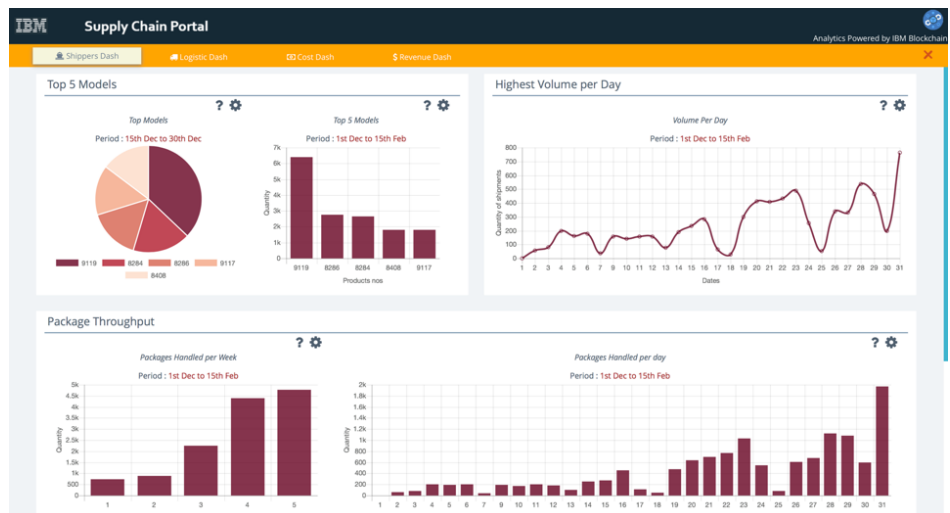

**Figure 2.** Analytics dashboard.

The analytics service consists of three main components namely, an authentication and dashboard management component, an analytics server, and ledger block reader. In our implementation and a typical deployment, all these components of the blockchain analytics service are co-deployed with one blockchain peer running on an IBM Blockchain Platform (IBP) [22]. This allows the service to operate directly on the blockchain data without ETL. However, the service can be deployed with more than one peer if needed. Also, it does not have any dependencies on IBP and it can be deployed with the peer running in any environment supported by Hyperledger Fabric.

The web-based user interface connects to an authentication and dashboard management component which takes care of user registration and authentication using their blockchain network credentials. This component also provides basic functions for dashboards configuration management.

The actual analytics is performed by the analytics server component. Users can access the analytics server component only after successful authentication. This component provides support for querying the blockchain data and additional analytics operations such as additional filtering, etc.

In order to enable analytics such as those depicted in Figure 2, the analytics server needs to be able to efficiently support a variety of aggregate queries. In Hyperledger Fabric blockchain solutions, data is stored as key-value pairs on the distributed ledger and, additionally, the latest value of each key is also stored in a so-called "state database" (implemented e.g. as CouchDB [26]) which facilitates efficient access to the latest version of data. Since the analytics service potentially needs to query the full history of all keys, data in the state database is not sufficient, and instead, all data on the ledger needs to be potentially queried. Unfortunately, in the current implementations of the ledger, complex and efficient queries are not supported. To overcome this problem, one of the key components of our analytics service, the ledger block reader, maintains a copy of historical ledger data in the same physical database where the state database resides. This way, we can leverage the query capabilities (of e.g.: CouchDB) which are significantly more powerful than the current ledger query capabilities. Our solution assumes that values of blockchain data are JSON documents.

While the described solution with its main components currently focuses primarily on descriptive analytics and queries, similar architecture and approach can be used for predictive and other advanced analytics as well. The advantage of the described approach is that the blockchain data does not need to leave the security zone of the blockchain solution and it is analyzed securely in the same environment.

## 7. Transparent Management of Data and Processing

Blockchain technologies are allowing a fresh look at how business processes can be designed, specified, implemented and maintained. The academic and industrial community is drawing on tried and true approaches to Business Process Management (BPM), such as activity-based process management (i.e.: the BPMN standard), the more data-centric case management approach (i.e.: the CMMN standard), and business rules and decision trees (i.e.: SBVR and DMN), respectively. The industry is adapting and extending those approaches to take advantage of fundamental characteristics of blockchain and to address emerging application use cases.

Blockchain provides a trusted data repository with privacy-preserving access controls. In the emerging blockchain-enabled business process solutions this is leveraged by focusing on constructs for asset and participant. Assets correspond to physical or conceptual business-relevant objects that may change as business processes evolve. Many blockchain-enabled solutions are centered around a family of asset types, including data that accumulates around them and possible lifecycle pathways that they may progress through. In applications relating to academic publishing, the asset types will focus on asset types such as articles, supporting datasets, reviews and editorial decisions, subscriptions, and access transactions. The business process modelling community is now drawing on academic work on business objects [27], business artifacts [28], and case management [29, 30] to develop explicit mechanisms to model, specify and implement the assets that underly collaborative processes, along with the possible lifecycles that they can traverse through. The two most used lifecycle models are based on finite state machines (and their generalization, state charts [31]) and directed acyclic graphs (DAGs) with rollback [32, 33]. Although the field is still evolving, the paradigm of Artifact-Centric Service Interoperation (ACSI) hubs [34, 35], which proposes the use of assets with lifecycles as the anchor for collaborative business processes, provides one blueprint for how blockchain solutions will be specified. Recent work on modeling and specifying constraints on the interaction of assets (e.g., [36, 37] will be an important extension of the ACSI hub approach, enabling enforceable controls on how collections of assets will work together.

In blockchain-supported business collaborations, there are multiple participants (including both individuals and organizations) that can access information stored on the hub. Various mechanisms are emerging to specify and enforce access controls. For example, Hyperledger Fabric currently supports channels, which involve a fixed set of participants who have the exclusive right to see data and invoke transactions within the channel. Another approach, supported by ACSI hubs [34] but not yet by blockchain frameworks, enables access controls at the level of assets and their attributes, e.g., so that an editor can see an article review and the reviewer's name, whereas the authors can see only the review. Importantly, participants can also access the smart contracts, or business logic, that is executing on the blockchain. The smart contracts can specify how assets progress along their lifecycle (e.g., when the final review of a paper has been submitted then forward to editor for a decision), and also constraints within or between assets (e.g., articles involving experiments must have all supporting data committed into the blockchain solution before reviewing can begin). The fact that the smart contracts can be viewed and agreed upon by the participants is central to enabling both the transparency and the auditability of all processing that occurs on the blockchain.

The use of blockchain can transform several of the core business processes that arise in academic publishing. Some of these processes are found in other domains, such as improved efficiency and reduced dispute resolution around invoicing, billing and payment; or such as enabling analytics about processes that span multiple organizations, while preserving individual privacy protections. Others are more specific to publishing, e.g., to create more transparent, trusted and auditable reviewing procedures, including the development of evidence-based evaluations of reviewer performance and expertise as an aid to editorial decisions. We focus here on two application areas that are of central interest to academic publishing. For both of these, a permissioned blockchain environment can provide a trusted, cost-efficient, sustainable approach that can provide important capabilities that are elusive with current practices.

The first application area relates to trusted verification of source data, algorithms, intermediate results, and published results. A blockchain network maintained by a family of leading international universities, industrial research labs, governmental research funding agencies, professional organizations, and publishing companies can provide a trusted, neutral hosting ground for this kind of "provenance" information, either for a single discipline or across disciplines. Not all data needs to be stored on the blockchain. Instead, some of the data can be stored off-chain but with associated cryptographic hashes stored on-chain; these can be used to verify that data has not subsequently been tampered with. The availability of full source data can be relevant across all disciplines. For example, it will streamline the reviewing process, enable follow-on research to have easier access to relevant source materials, and help to overcome fallacious references to non-existent or non-relevant sources. Availability of provenance information for papers in the experimental sciences will be especially relevant in reducing falsification of data and/or inferred results. A central component of maintaining this trusted provenance information will be the transparency of the processes used to enforce the gathering and maintenance of the data. The general accessibility of smart contract logic to all participants will be a key enabler for this transparency.

A second key application area of blockchain for academic publishing concerns the development of transparent, systematic approaches for evaluating the academic quality and impact of papers and publishing outlets. At present Google Scholar is a primary source of citation indexes, which have become a proxy for measuring the scientific, academic, and scholarly impact of both publications and the researchers who write them. But there is little visibility into the algorithms that underlie Google Scholar, nor much guarantee that all of the author's publications are included. Further, there is at present no attempt to enable a focus on citations from more reputable publication venues and reduce or eliminate the influence of predatory publication outlets. A blockchain-enabled solution, maintained by a consortium of leading universities and governmental research agencies, could provide a transparent, systematic framework and family of processes and algorithms for maintaining data about all publications and publication outlets. Multiple algorithms could be developed and made available for tallying the academic merit of articles, authors, and publication outlets. The algorithms would be accessible as smart contracts, so that interested stakeholders could understand how importance was measured, including how publications outlets individual citations were weighted. In particular, information about the reviewing and possible pay-for-publishing procedures of different publication outlets could be incorporated into the weighting functions in a transparent manner. The availability of a trusted blockchain-enabled solution to provide this kind of informed measure of academic merit could bring important benefits to the scientific community in the form of more accurate assessment of research impact, and both editorial decisions and research funding decisions that are better informed.

## 8. Personal Data Management on Blockchain

In this section, we look at how blockchain supports the management of the ownership of personal data. First, consider a publication submission process implemented on blockchain. As data is collected in the blockchain around a publication submission, the process continues to enrich that data with assigned reviewers, reviews, addendums, decisions, and ultimately the camera-ready publication. The process will enhance transparency where needed and enforce privacy when in need to protect the data. But ultimately much of that data belongs to the author, for his or her own consumption, analysis, or for simple compliance reasons.

The personal data collected by the blockchain enabled processes require cautious management. Recently the Global Data Privacy Regulation, better known as GDPR [38] came to effect in the European Union. The regulation mandates that every user is entitled to his or her own data. It is possible then for anyone to collect all the data about him or her and require a medium to store that information. For example, the Personal Data Store (PDS) [39] is a storage mechanism to collect, protect, and manage personal data. The owner can grant or restrict access to specific data items. Zyskind et.al., [40] propose to use the Blockchain to act as an automated access-control manager that does not require trust from a third party.

Owning the personal data creates a new paradigm, the user can selectively combine data items to create an intent. By intent, we mean the desire to act combined with the necessary data to support that intent. Let's take an example from Academic Publishing: the selection of a program committee and how using intents we can create better matches. A program committee is usually selected a priori with limited understanding of the skills or even the preferences or areas of interest of each member. What if each potential member expresses his or her areas of interest, supports it with related work, or ratings gathered as a reviewer, and even expresses the level of availability to participate? Conceivably, a potential Program Committee member may express his unsolicited intent to participate allowing to uncover other untapped talents. At that point, a Program Chair can better understand what are the areas covered, seek other members to fill gaps or fine tune the Call for Papers to address the true interests of the program committee. Intents can also be used when assigning publications, to fine-tune a particular area of specialty or potentially combined with the external blockchains to detect conflicts of interest based on prior co-authorships or collaborations.

Finally, by expressing an intent the user can be more precise on its action plan, as opposed to reacting to a solicitation that may be broad in nature, and with the support of a PDS, it may be possible to furnish necessary supporting facts to properly respond to that intent by the fulfilling party.

## 9. Summary

We have presented permissioned blockchain technologies and outlined how they can be leveraged in the domain of academic publishing. We have explained the differences between public and private blockchains as well as the key concepts such as consensus and security considerations. We have also presented the latest technologies and tools available including IBM Blockchain Platform which allows to efficiently build and operate cloud-based blockchain solutions, and Secure Blockchain Analytics which allow to tap into the value of the data stored on blockchain ledgers. We have described how blockchains provide trust and facilitate collaboration without the need for centralized management and authority, something that the academic publishing, centered on information processing and reliant on interactions between globally distributed participants will greatly benefit from. As such, blockchains provide the opportunity and building blocks to design new solutions, which will address shortcomings of the present technologies as well as significantly transform and extend the existing processes and modes of operation to the benefit of the academic publishing community.